\documentclass[11pt]{article}

\usepackage{amsmath}
\usepackage{bm}
\usepackage{amssymb}
\usepackage{amstext}
\usepackage{latexsym}
\usepackage{color}
\usepackage{graphicx}
\usepackage{epsfig}
\usepackage{subfigure}
\usepackage{rotating}
\usepackage{curves}
\usepackage{epic}
\usepackage{amsthm}
\allowdisplaybreaks[1]

\usepackage{amsfonts}
\usepackage[T1]{fontenc}
\usepackage[latin1]{inputenc}
\usepackage{authblk}
\usepackage{tikz}
\usetikzlibrary{shapes,arrows}
\usepackage{graphics}
\usepackage{verbatim}
\usepackage{color}
\usepackage{hyperref}
\usepackage{mathtools}

\newcommand{\mb}{\mathbb}
\newcommand{\mc}{\mathcal}
\newcommand{\be}{\begin{equation}}
\newcommand{\en}{\end{equation}}
\newcommand{\id}{{\mb I}}

\newtheorem{defi}{Definition}[section]
\newtheorem{lem}[defi]{Lemma}

\newtheorem{Theo}{Theorem}[section]

\newtheorem{Prop}[Theo]{Proposition}

\newtheorem{remark}[Theo]{Remark}

\newcommand{\bedefin}{\begin{defi}}
\newcommand{\findefi}{\end{defi} \medskip}

\newcommand{\belem}{\begin{lem}$\!\!${\bf }}
\newcommand{\enlem}{\end{lem}}

\newcommand{\prf}{\noindent{\bf{ Proof}\,\,}}

\newcommand{\beno}{\begin{equation*}}
\newcommand{\enno}{\end{equation*}}

\newcommand{\bea}{\begin{eqnarray}}
\newcommand{\ena}{\end{eqnarray}}

\newcommand{\nn}{\nonumber}

\newcommand{\bxi}{\mbox{\boldmath $\xi$}}

\newcommand\NC{\text{nc}}

\newcommand{\bp}{\mathbf p}
\newcommand{\bq}{\mathbf q}
\newcommand{\bz}{\mathbf z}
\newcommand{\bX}{\mathbf X}
\newcommand{\br}{\mathbf r}
\newcommand{\nchi}{\chi_{\mathbf{q},\mathbf{p}}^{\NC}}
\newcommand{\nnchi}{\tilde{\chi}_{\mathbf{q},\mathbf{p}}^{\NC}}
\newcommand{\neta}{\eta_{\mathbf{q},\mathbf{p}}^{\NC}}
\newcommand{\nneta}{\tilde{\eta}_{\mathbf{q},\mathbf{p}}^{\NC}}
\newcommand{\nnneta}{\tilde{\eta}_{\mathbf{q}^{\prime},\mathbf{p}^{\prime}}^{\NC}}
\newcommand{\netap}{\eta_{\mathbf{q}^{\prime},\mathbf{p}^{\prime}}^{\NC}}
\newcommand{\g}{G_{\NC}}
\newcommand{\nQ}{q_{1}^{\NC}}
\newcommand{\nR}{q_{2}^{\NC}}
\newcommand{\nS}{p_{1}^{\NC}}
\newcommand{\nT}{p_{2}^{\NC}}

\newcommand\CC{\mathbf C}
\newcommand\RR{\mathbf R}
\newcommand\FFh{\mathcal F_{\hbar}}
\newcommand\WW{\mathcal W}
\newcommand\cB{\mathcal B}
\newcommand\spr[1]{\langle#1\rangle}
\newcommand\bznc{\bz^{\NC}}
\newcommand\bZnc{\bz^{\prime\NC}}
\newcommand\cT{\mathcal T}
\newcommand\cP{\mathcal P}
\newcommand\cC{\mathcal C}
\newcommand\vthta{\vartheta}

\newcommand\GWH{G_{\text{\tiny{WH}}}}
\newcommand\tFFh{\tilde\FFh}
\newcommand\tWW{\tilde{\WW}}
\newcommand\znc{z^{\NC}}
\newcommand\Znc{z^{\prime\NC}}
\newcommand\tcT{\tilde{\cT}}
\newcommand\tcC{\tilde{\cC}}
\newcommand\tV{\tilde V}
\newcommand\tnu{\tilde\nu}

\catcode `\@=11 \@addtoreset{equation}{section}

\catcode `\@=12

\begin{document}

\title{Noncommutative coherent states and related aspects of Berezin-Toeplitz quantization}
\author[1]{S. Hasibul Hassan Chowdhury\thanks{shhchowdhury@gmail.com}}
\author[2]{S. Twareque Ali\thanks{twareque.ali@concordia.ca}}
\author[3,4]{Miroslav Engli\v{s}\thanks{englis@math.cas.cz}}
\affil[1]{Chern Institute of Mathematics, Nankai University, Tianjin 300071, China}
\affil[2]{Department of Mathematics and Statistics, Concordia University, Montr\'eal, Qu\'ebec, Canada H3G 1M8}
\affil[3]{Mathematics Institute, Silesian University in Opava, Na~Rybn\'{\i}\v{c}ku~1, 74601 Opava, Czech Republic}
\affil[4]{Mathematics Institute, Czech Academy of Sciences, \v{Z}itn\'a~25, 11567~Prague~1, Czech Republic}
\date{\today\\
\qquad\\
\it Dedicated by the first and the third authors to the memory of the second author,\\
with gratitude for his friendship and for all they learnt from~him}

\maketitle

\begin{abstract}
In~this paper, we construct noncommutative coherent states using various families of unitary irreducible representations (UIRs) of $\g$, a connected, simply connected nilpotent Lie group, that was identified as the kinematical symmetry group of noncommutative quantum mechanics for a system of 2-degrees of freedom in an earlier paper.
Likewise described are the degenerate noncommutative coherent states arising from the degenerate UIRs of $\g$.
We~then compute the reproducing kernels associated with both these families of coherent states and study Berezin-Toeplitz quantization of the observables on the underlying 4-dimensional phase space,
analyzing in particular the semi-classical asymptotics for both these cases.
\end{abstract}

\section{Introduction}\label{sec:intro}
Noncommutative quantum mechanics (NCQM) is an active field of research these
days. The starting point here is to alter the canonical commutation relations
(CCR) among the respective positions and momenta coordinates and hence
introduce a new noncommutative Lie algebra structure. Consult
\cite{scholtzt,delducetalt} for a detailed account on this approach.

There is another approach available where one starts with noncommutative field theory (NCFT) studying quantum field theory on noncommutative space-time. An excellent pedagogical treatment to noncommutative quantum field theory can be found in \cite{Szabo}. Refer to the excellent review \cite{Douglasetal} to delve further into the studies of NCFT. A noncommutative quantum field theory is the one where the fields are functions of space-time coordinates with spatial coordinates failing to commute with each other. Among others, Snyder and Yang were the leading proponents to introduce the concept of noncommutative structure of space-time (see \cite{Sny,Yang}). Introduction of such assumption of spatial noncommutativity eliminates ultraviolet (UV) divergences of quantum field theory and runs parallel to the technique of renormalization as a cure to such annoying divergence issues in quantum field theory. NCQM can then be seen as nonrelativistic approximation of NCFT (see \cite{Hoetal,Chaichianetal}).

In yet another approach (see \cite{ncqmjmp},\cite{ncqmjpa}), the authors start from a certain nilpotent Lie group $\g$ as the defining group of NCQM for a system of 2 degrees of freedom and compute its unitary dual using the method of orbits introduced by Kirillov (see \cite{Kirillovbook}). Although $\g$ does not contain the Weyl-Heisenberg group $\GWH$ as its subgroup, the unitary dual of $\GWH$ is found to be sitting inside that of $\g$. Various gauges arising in NCQM are also found to be related to a certain family of unitary irreducible representations (UIRs) of $\g$.  The Lie group $\g$ was later identified as the kinematical symmetry group of NCQM in \cite {wigjmp} where various Wigner functions for such model of NCQM were computed that were found to be supported on relevant families of coadjoint orbits associated with $\g$. The purpose of this paper is to construct coherent states arising from $\g$, study their properties and discuss the associated Berezin-Toeplitz quantization of the observables on the 4-dimensional phase space, including the relevant ``semi-classical'' asymptotics.

In~Section~\ref{sec:cs ncqm}, we~review the relevant facts about the UIRs of
the 7-dimensional real Lie group $\g$ from \cite{ncqmjpa} and proceed to
construct the associated coherent states and reproducing kernel.
The~UIRs in question are parameterized by triples $(\hbar,\vthta,\cB)\in\RR^3$
satisfying $\hbar^2-\vthta\cB\neq0$;
the ``degenerate'' case $\hbar^2=\vthta\cB$ is then discussed separately in
Section~\ref{sec:degenrt-rep}.
With~the coherents states in hand, one~constructs the Toeplitz operators in
the standard fashion, and we~examine the corresponding ``semiclassical limit''
of the resulting Berezin-Toeplitz quantization in Section~\ref{secTOEP}.
The~main novelty here, of~course, is~the presence of the three deformation
parameters $\hbar,\vthta,\cB$ instead of the sole Planck constant~$\hbar$,
thus making it already somewhat unclear in what manner these three should be
allowed to approach zero in any analogue of the ordinary semi-classical limit
where one just has $\hbar\searrow0$; note that $(\hbar,\vthta,\cB)$ cannot
approach $(0,0,0)$ completely unrestrictedly due to the non-degenaracy
condition $\hbar^2-\vthta\cB\neq0$. It~turns out that the right objects from
this point of view are the renormalized quantities $B:=\cB/\hbar$ and
$T:=\vthta/\hbar$, and nice semiclassical asymptotics are established for
the situation when $\hbar,B,T$ all tend to zero, without any restrictions.
(Note that $B$ has the physical interpretation of the applied magnetic field,
see eqn.~(3.6) in \cite{ncqmjpa}. Note also that in the $B,T$ notation the
non-degeneracy condition becomes simply $BT\neq1$, so~there is no longer
any problem with $B,T,\hbar$ simultaneously approaching zero.) 

One~can in principle consider the Berezin-Toeplitz quantization and the
corresponding semi-classical behaviour also in the degenerate case
$\hbar^2-\vthta\cB=0$. In~this case, the~underlying Hilbert space
carrying the coherent states undergoes a dimension reduction from 4 to~2,
and while we still get a unique associated reproducing kernel,
it~turns out that the measure defining the inner product in the reproducing
kernel Hilbert space at hand is no longer uniquely determined.
Furthermore, the~semi-classical asymptotics of the associated Berezin-Toeplitz
operators turn out not to depend at all on the choice of this measure,
nor in fact on the deformation parameter~$\vthta$ (or~on~$\cB=\hbar^2/\vthta$),
and reduce just to the plain Berezin-Toeplitz deformation quantization 
(star-product) on the complex plane.
Details are supplied in Section~\ref{secTOEPdeg}.

\section*{Acknowledgements} The~third author (M.E) was supported by GA~\v{C}R
grant no.~201/12/G028 and Institutional funding (RVO) for I\v{C}~47813059.
The~first author (S.~H.~H.~C.) gratefully acknowledges a grant from National
Natural Science Foundation of China (NSFC) under Grant No.~11550110186.

The~middle named author, S.~Twareque~Ali, passed away unexpectedly on
January~25 2016, before this manuscript was finished.
The~other two authors dedicate this paper to~him, in~memory of his scientific
enthusiasm and great and lasting friendship.

\section{Noncommutative coherent states associated to $\g$}
\label{sec:cs ncqm}
The geometry of the coadjoint orbits associated with the 7-dimensional real Lie group $\g$ is studied in detail in \cite{ncqmjpa}. There the orbits were classified based on the triple $(\rho,\sigma,\tau)\in\mathbb{R}^{3}$. In this paper, we will focus on the family of UIRs corresponding to $(\rho,\sigma,\tau)\in\mathbb{R}^{3}$ such that $\rho\neq 0$, $\sigma\neq 0$ and $\tau\neq 0$ with $\rho^{2}\alpha^{2}-\gamma\beta\sigma\tau\neq 0$. The UIRs are given by
\bea\label{eq:UIR-gen-NCQM}
\lefteqn{(\hat{U}^{\rho}_{\sigma,\tau}(\theta,\phi,\psi,\bq,\bp)f)(r_{1},s_{2})}\nonumber\\
    &&=e^{i\rho(\theta+\alpha q_{2}s_{2}+\alpha p_{1}r_{1}+\frac{\alpha}{2}q_{1}p_{1}-\frac{\alpha}{2}q_{2}p_{2})}e^{i\sigma(\phi+\beta p_{1}s_{2}-\frac{\beta}{2}p_{1}p_{2})}\nonumber\\
    &&\times e^{i\tau(\psi+\gamma q_{2}r_{1}+\frac{\gamma}{2}q_{2}q_{1})}f(r_{1}+q_{1},s_{2}-p_{2}).
\ena
Note that in (\ref{eq:UIR-gen-NCQM}), $r_{1}$ has the dimension of length and $s_{2}$ has that of momentum. By taking the inverse Fourier transform of (\ref{eq:UIR-gen-NCQM}) in the second coordinate $s_2$ yields the following representation on $L^{2}(\mathbb{R}^{2},d\br)$:
\bea\label{eq:UIR-Innv-FT-gen-NCQM}
\lefteqn{(U^{\rho}_{\sigma,\tau}(\theta,\phi,\psi,\bq,\bp)f)(r_{1},r_{2})}\nonumber\\
    &&=e^{i\rho(\theta+\alpha p_{1}r_{1}+\alpha p_{2}r_{2}+\frac{\alpha}{2}q_{1}p_{1}+\frac{\alpha}{2}q_{2}p_{2})}e^{i\sigma(\phi+\frac{\beta}{2}p_{1}p_{2})}\nonumber\\
    &&\times e^{i\tau(\psi+\gamma q_{2}r_{1}+\frac{\gamma}{2}q_{1}q_{2})}f\left(r_{1}+q_{1},r_{2}+q_{2}+\frac{\sigma\beta}{\rho\alpha}p_{1}\right),
\ena
where $f\in L^{2}(\mathbb{R}^{2},d\br)$.

\bedefin\label{cs-def}
For a given fixed vector $\chi\in L^{2}(\mathbb{R}^{2},d\br)$ and a fixed point $(\bq,\bp)\in\mathbb{R}^{4}$, the underlying phase space for the 2-dimensional system under study, let us define the following vectors in $L^{2}(\mathbb{R}^{2},d\br)$ as
\be\label{def-nc-cs}
\nchi=U^{\rho}_{\sigma,\tau}(0,0,0,-\bq,\bp)\chi.
\en
Introducing the following change of variables:
\beno
\hbar=\frac{1}{\rho\alpha},\;\;\vartheta=-\frac{\sigma\beta}{\rho^{2}\alpha^{2}}\;\hbox{and}\;\mathcal{B}=-\frac{\tau\gamma}{\rho^{2}\alpha^{2}},
\enno
the vectors $\nchi$ read
\beno
\nchi(\br)=e^{\frac{i}{\hbar}\left(\br-\frac{1}{2}\bq\right).\bp-\frac{i\vartheta}{2\hbar^{2}}p_{1}p_{2}+\frac{i\mathcal{B}}{\hbar^{2}}\left(q_{2}r_{1}-\frac{1}{2}q_{1}q_{2}\right)}\chi\left(r_{1}-q_{1},r_{2}-q_{2}-\frac{\vartheta}{\hbar}p_{1}\right).
\enno
Let us now define the noncommutative coherent states using the vectors $\nchi$ as
\be \label{normlzd-nc-cs}
\neta(\br)=e^{\frac{i}{\hbar}\left(\br-\frac{1}{2}\bq\right).\bp-\frac{i\vartheta}{2\hbar^{2}}p_{1}p_{2}+\frac{i\mathcal{B}}{\hbar^{2}}\left(q_{2}r_{1}-\frac{1}{2}q_{1}q_{2}\right)}\eta\left(r_{1}-q_{1},r_{2}-q_{2}-\frac{\vartheta}{\hbar}p_{1}\right),
\en
where $\eta$ is a vector given by $\eta=\frac{\chi}{\vert\vert\chi\vert\vert}$.
\findefi
The phase space for the 2-dimensional noncommutative system is
$\mathbb{R}^{4}$. The underlying observables are functions defined over the
phase space variables $\bq$, $\bp$. These functions are taken to be elements of
the Hilbert space $L^{2}(\mathbb{R}^{4},d\nu(\bq,\bp))$ equipped with the
measure
$d\nu(\bq,\bp)=\frac{\vert\hbar^{2}-\mathcal{B}\vartheta\vert}
 {4\pi^{2}\hbar^{4}}d\bq\;d\bp$.
The noncommutative coherent states $\neta$, given by (\ref{normlzd-nc-cs}),
satisfy the resolution of identity as stated by the following lemma: 
\belem\label{lem-res-idnty}
The vectors $\neta$ defined as the noncommutative coherent states by (\ref{normlzd-nc-cs}) satisfy the following integral relation:
\be\label{stmnt-res-idnty}
\int_{\mathbb{R}^{4}}\vert\neta\rangle\langle\neta\vert\;d\nu(\bq,\bp)=\mathbb{I},
\en
where $\mathbb{I}$ is the identity operator on $L^{2}(\mathbb{R}^{2},d\br)$.
\enlem
\prf{.}
Introducing the following changes of variables:
\be\label{nc-chng-vrbl}
\begin{aligned}
q_{1}^{\NC}&=q_{1}\\
q_{2}^{\NC}&=q_{2}+\frac{\vartheta}{\hbar}p_{1}\\
p_{1}^{\NC}&=p_{1}+\frac{\mathcal{B}}{\hbar}q_{2}\\
p_{2}^{\NC}&=p_{2},
\end{aligned}
\en
the noncommutative coherent states $\neta$, appearing in (\ref{normlzd-nc-cs}), can neatly be written as
\be\label{nc-cs-smpfd}
\neta(\br)=e^{\frac{i}{\hbar}\left(\br-\frac{1}{2}\bq^{\NC}\right).\bp^{\NC}}\eta\left(\br-\bq^{\NC}\right),
\en
where $(q_{1}^{\NC},q_{2}^{\NC})$ and $(p_{1}^{\NC},p_{2}^{\NC})$ are denoted as $\bq^{\NC}$ and $\bp^{\NC}$, respectively.

The associated measures transform as
\be\label{jacbn-ncqm-qm}
d\bq^{\NC}d\bp^{\NC}=\frac{\vert\hbar^{2}-\mathcal{B}\vartheta\vert}{\hbar^{2}}d\bq\;d\bp.
\en

Let us now choose two compactly supported smooth functions $f$ and $g$ in $L^{2}(\mathbb{R}^{2},d\br)$. One then obtains,
\bea\label{prf-res-idnty}
\lefteqn{\int_{\mathbb{R}^{4}}\langle f\vert\neta\rangle\langle\neta\vert g\rangle d\nu(\bq,\bp)}\nn\\
&&=\frac{\vert\hbar^{2}-\mathcal{B}\vartheta\vert}{4\pi^{2}\hbar^{4}}\int_{\mathbb{R}^{4}}\langle f\vert\neta\rangle\langle\neta\vert g\rangle d\bq\;d\bp\nn\\
&&=\frac{1}{4\pi^{2}\hbar^{2}}\int_{\mathbb{R}^{4}}\langle f\vert\neta\rangle\langle\neta\vert g\rangle d\bq^{\NC}d\bp^{\NC}\nn\\
&&=\frac{1}{4\pi^{2}}\int_{\mathbb{R}^{4}}\left[\int_{\mathbb{R}^{2}}\int_{\mathbb{R}^{2}}e^{\frac{i}{\hbar}(\br-\br^{\prime}).\bp^{\NC}}\overline{f(\br)}\;\eta(\br-\bq^{\NC})\overline{\eta(\br^{\prime}-\bq^{\NC})}g(\br^{\prime})\;d\br\;d\br^{\prime}\right]d\bq^{\NC}d\left(\frac{\bp^{\NC}}{\hbar}\right)\nn\\
&&=\int_{\mathbb{R}^{2}}\left[\int_{\mathbb{R}^{2}}\int_{\mathbb{R}^{2}}\delta^{(2)}(\br-\br^{\prime})\overline{f(\br)}\;\eta(\br-\bq^{\NC})\overline{\eta(\br^{\prime}-\bq^{\NC})}g(\br^{\prime})\;d\br\;d\br^{\prime}\right]d\bq^{\NC}\nn\\
&&=\int_{\mathbb{R}^{2}}\int_{\mathbb{R}^{2}}\overline{f(\br)}\;\eta(\br-\bq^{\NC})\overline{\eta(\br-\bq^{\NC})}g(\br)\;d\br\;d\bq^{\NC}\nn\\
&&=\vert\vert\eta\vert\vert^{2}\langle f\vert g\rangle\nn\\
&&=\langle f\vert g\rangle.
\ena
Using the continuity of the inner product of the underlying Hilbert space and the fact that the compactly supported smooth functions are dense in $L^{2}(\mathbb{R}^{2},d\br)$, one can extend the above equality to any pair of functions $f$, $g$ in $L^{2}(\mathbb{R}^{2},d\br)$ and hence proving the lemma.
\qed

Let us rewrite (\ref{nc-cs-smpfd}) and observe that
\bea\label{cs-smplfd-oprtorial}
\neta(\br)&=&e^{\frac{i}{\hbar}\left(\br-\frac{1}{2}\bq^{\NC}\right).\bp^{\NC}}\eta\left(\br-\bq^{\NC}\right)\nn\\
&=&e^{\frac{i}{\hbar}\left(\br-\frac{1}{2}\bq^{\NC}\right).\bp^{\NC}}e^{\left(-q^{\NC}_{1}\frac{\partial}{\partial r_{1}}-q^{\NC}_{2}\frac{\partial}{\partial r_{2}}\right)}\eta(\br)\nn\\
&=&e^{\frac{i}{\hbar}\left[\br.\bp^{\NC}-q^{\NC}_{1}\left(-i\hbar\frac{\partial}{\partial r_{1}}\right)-q^{\NC}_{2}\left(-i\hbar\frac{\partial}{\partial r_{2}}\right)\right]}\eta(\br)\nn\\
&=&e^{\frac{i}{\hbar}(\bxi^{T}\omega^{\NC}\bX)}\eta(\br),
\ena
where $\bxi$ and $\bX$ are $4\times 1$ column vectors given by
\be\label{nc-dis-op-symbls}
\bxi=\begin{bmatrix}q^{\NC}_{1}\\q^{\NC}_{2}\\p^{\NC}_{1}\\p^{\NC}_{2}\end{bmatrix}=\begin{bmatrix}q_{1}\\q_{2}+\frac{\vartheta}{\hbar}p_{1}\\p_{1}+\frac{\mathcal{B}}{\hbar}q_{2}\\p_{2}\end{bmatrix}\;\;\hbox{and}\;\;
\bX=\begin{bmatrix}\hat{Q}^{\NC}_{1}\\\hat{Q}^{\NC}_{2}\\\hat{P}^{\NC}_{1}\\\hat{P}^{\NC}_{2}\end{bmatrix}=\begin{bmatrix}r_{1}+i\vartheta\frac{\partial}{\partial r_{2}}\\r_{2}\\-i\hbar\frac{\partial}{\partial r_{1}}\\-\frac{\mathcal{B}}{\hbar}r_{1}-i\hbar\frac{\partial}{\partial r_{2}}\end{bmatrix},
\en
while $\omega^{\NC}$ in (\ref{cs-smplfd-oprtorial}) is given by the following $4\times 4$ matrix:
\be\label{nc-anlg-J}
\omega^{\NC}=\begin{bmatrix}0&0&-1&0\\-\frac{\mathcal{B}\hbar}{(\hbar^{2}-\mathcal{B}\vartheta)}&0&0&-\frac{\hbar^{2}}{(\hbar^{2}-\mathcal{B}\vartheta)}\\\frac{\hbar^{2}}{\hbar^{2}-\mathcal{B}\vartheta}&0&0&\frac{\vartheta\hbar}{\hbar^{2}-\mathcal{B}\vartheta}\\0&1&0&0\end{bmatrix}.
\en
The 4 entries of the column vector $\bX$, i.e. $\hat{Q}^{\NC}_{i}$, $\hat{P}^{\NC}_{i}$ for $i=1,2$,  in (\ref{nc-dis-op-symbls}) represent the non-central generators of $\g$ in the Landau gauge representation defined on $L^{2}(\mathbb{R}^{2},d\br)$ as obtained in (\cite{ncqmjpa}).
\bedefin\label{nc-displcmnt-oprt-def}
The unitary operator $\mathcal{D}^{\NC}(\bq,\bp)$ that generates the noncommutative coherent state vectors $\neta$ by acting upon the normalized ground state vector $\eta$ is defined as the noncommutative displacement operator.
\bea\label{eq-nc-displc-opt-def}
\neta(\br)&=&\mathcal{D}^{\NC}(\bq,\bp)\eta(\br)\nn\\
&=&e^{\frac{i}{\hbar}\left(\bxi^{T}\omega^{\NC}\bX\right)}\eta(\br),
\ena
where $\bxi$, $\bX$ and $\omega^{\NC}$ are as given in (\ref{nc-dis-op-symbls}) and (\ref{nc-anlg-J}).
\findefi
\begin{remark}
A few remarks on the quantum mechanical limits of the above formulations are in order. From (\ref{nc-dis-op-symbls}) and (\ref{nc-anlg-J}), it can easily be seen that as $\vartheta\rightarrow 0$, $\mathcal{B}\rightarrow 0$, the noncommutative displacement operator $\mathcal{D}^{\NC}(\bq,\bp)$ approaches the canonical displacement operator:
\bea\label{lim-nc-displcmnt-oprtr}
\mathcal{D}^{\NC}(\bq,\bp)&\xrightarrow{\vartheta\rightarrow 0,\;\vartheta\rightarrow 0}&\exp\frac{i}{\hbar}\left(\begin{bmatrix}q_{1}&q_{2}&p_{1}&p_{2}\end{bmatrix}\begin{bmatrix}0&0&-1&0\\0&0&0&-1\\1&0&0&0\\0&1&0&0\end{bmatrix}\begin{bmatrix}\hat{Q}_{1}\\\hat{Q}_{2}\\\hat{P}_{1}\\\hat{P}_{2}\end{bmatrix}\right)\nn\\
&=&\exp\frac{i}{\hbar}(\bp.\mathbf{\hat{Q}}-\bq.\mathbf{\hat{P}}),
\ena
where $\mathbf{\hat{Q}}=(\hat{Q}_{1},\hat{Q}_{2})=(r_{1},r_{2})$ and $\mathbf{\hat{P}}=(\hat{P}_{1},\hat{P}_{2})=(-i\hbar\frac{\partial}{\partial r_{1}},-i\hbar\frac{\partial}{\partial r_{2}})$ are the standard quantum mechanical representation of the position and momentum operators of the underlying 2-dimensional system defined on $L^{2}(\mathbb{R}^{2},d\br)$. The cases $\mathcal{B}\rightarrow 0$ or $\vartheta\rightarrow 0$ can similarly be studied. For example,
\be\label{case-B-0}
\omega^{\NC}\xrightarrow{\mathcal{B}\rightarrow 0}\begin{bmatrix}0&0&-1&0\\0&0&0&-1\\1&0&0&\frac{\vartheta}{\hbar}\\0&1&0&0\end{bmatrix},\;\omega^{\NC}\xrightarrow{\vartheta\rightarrow 0}\begin{bmatrix}0&0&-1&0\\-\frac{\mathcal{B}}{\hbar}&0&0&-1\\1&0&0&0\\0&1&0&0\end{bmatrix}.
\en
\end{remark}

Now in view of (\ref{cs-smplfd-oprtorial}), the noncommutative coherent states $\neta$ can be read off as
\be\label{nc-cs-rep-ker}
\neta(\br)=e^{\frac{i}{\hbar}(p_{1}^{\NC}\hat{Q}_{1}+p_{2}^{\NC}\hat{Q}_{2}-q_{1}^{\NC}\hat{P}_{1}-q_{2}^{\NC}\hat{P}_{2})}\eta(\br).
\en
Let us choose an element $\phi\in L^{2}(\mathbb{R}^{2},d\br)$ and consider the map $\Phi:\mathbb{R}^{4}\rightarrow\mathbb{C}$ for a fixed vector $\phi\in L^{2}(\mathbb{R}^{2},d\br)$ by
\be\label{def-map}
\Phi(\bq,\bp)=\langle\neta\vert\phi\rangle.
\en
It is immediate that $\Phi\in L^{2}(\mathbb{R}^{4},d\nu(\bq,\bp))$. Now consider the isometry map
\be\label{iso-map-def}
W:L^{2}(\mathbb{R}^{2},d\br)\rightarrow L^{2}(\mathbb{R}^{4},d\nu)
\en
defined by $W\phi=\Phi$. The range $\mc W$ of this isometry map, a~closed subspace of $L^{2}(\mathbb{R}^{4},d\nu)$, is~a~reproducing kernel Hilbert space (RKHS) and the function $K^{\NC}:\mathbb{R}^{4}\times\mathbb{R}^{4}\rightarrow \mathbb{C}$ with
\be\label{rep-ker-def}
K^{\NC}(\bq,\bp;\bq^{\prime},\bp^{\prime})=\langle\neta\vert\netap\rangle
\en
is its reproducing kernel.

Let us now recall a few facts about $\g$ from \cite{ncqmjmp}. The group composition rule for $\g$ are given by
\bea\label{grplawtriplyextendedgrp}
\lefteqn{(\theta,\phi,\psi,\bq,\bp)(\theta^{\prime},\phi^{\prime},\psi^{\prime},\bq^{\prime},\bp^{\prime})}\nonumber\\
&&=(\theta+\theta^{\prime}+\alpha\;\xi((\bq,\bp),(\bq^{\prime},\bp^{\prime})),\phi+\phi^{\prime}+\beta\;\xi^{\prime}((\bq,\bp),(\bq^{\prime},\bp^{\prime}))\nn\\
&&\;\;\;,\psi+\psi^{\prime}+\gamma\;\xi^{\prime\prime}((\bq,\bp),(\bq^{\prime},\bp^{\prime})),\bq+\bq^{\prime},\bp+\bp^{\prime}),
\ena
where the 3-inequivalent local exponents of the abelian group of translations in $\mathbb{R}^{4}$ is given by
\be\label{local-expnts}
\begin{aligned}
\xi((\bq,\bp),(\bq^{\prime},\bp^{\prime}))&=\frac{1}{2}[q_{1}p_{1}^{\prime}+q_{2}p_{2}^{\prime}-p_{1}q_{1}^{\prime}-p_{2}q_{2}^{\prime}],\\
\xi^{\prime}((\bq,\bp),(\bq^{\prime},\bp^{\prime}))&=\frac{1}{2}[p_{1}p_{2}^{\prime}-p_{2}p_{1}^{\prime}],\\
\xi^{\prime\prime}((\bq,\bp),(\bq^{\prime},\bp^{\prime}))&=\frac{1}{2}[q_{1}q_{2}^{\prime}-q_{2}q_{1}^{\prime}].
\end{aligned}
\en
\begin{Prop}
Provided one chooses the ground state vector $\eta$ in (\ref{nc-cs-rep-ker}) to be the following normalized Gaussian function:
\be\label{grnd-state-gaussn}
\eta(\br)=\frac{1}{\sqrt{\pi}s}e^{-\frac{1}{2s^{2}}\vert\br\vert^{2}},
\en
then, the reproducing kernel $K^{\NC}$ (see \ref{rep-ker-def}) associated with the Lie group $\g$ reads
\bea\label{rep-krnl-dfn-grp-ncqm}
\lefteqn{K^{\NC}((\bq,\bp),(\bq^{\prime},\bp^{\prime}))}\nn\\
&&=e^{\frac{i}{\hbar}\xi((\bq,\bp),(\bq^{\prime},\bp^{\prime}))-\frac{s^{2}}{4\hbar^{2}}\vert\bp-\bp^{\prime}\vert^{2}-\frac{1}{4s^{2}}\vert\bq-\bq^{\prime}\vert^{2}}e^{\frac{i\vartheta}{\hbar^{2}}\xi^{\prime}((\bq,\bp),(\bq^{\prime},\bp^{\prime}))}\nn\\
&&\times e^{-\frac{1}{4s^{2}}\left[\frac{\vartheta^{2}}{\hbar^{2}}(p_{1}-p_{1}^{\prime})^{2}+2\frac{\vartheta}{\hbar}(q_{2}-q_{2}^{\prime})(p_{1}-p_{1}^{\prime})\right]}e^{\frac{i\mathcal{B}}{\hbar^{2}}\xi^{\prime\prime}((\bq,\bp),(\bq^{\prime},\bp^{\prime}))-\frac{s^{2}}{4\hbar^{2}}\left[\frac{\mathcal{B}^{2}}{\hbar^{2}}(q_{2}-q_{2}^{\prime})^{2}+2\frac{\mathcal{B}}{\hbar}(p_{1}-p_{1}^{\prime})(q_{2}-q_{2}^{\prime})\right]},\nn\\
\ena
where $\xi$, $\xi^{\prime}$ and $\xi^{\prime\prime}$ are all given by (\ref{local-expnts}). Also, $s$ stands for the standard deviation associated with the position vector $\br=(r_1,r_2)$ and hence has the dimension of length.
\end{Prop}
\prf{.}
Using (\ref{nc-cs-rep-ker}), one finds that
\bea\label{rep-ker-proof}
\lefteqn{\langle\neta\vert\netap\rangle}\nn\\
&&=\langle\eta\vert e^{-\frac{i}{\hbar}(\bp^{\NC}.\mathbf{\hat{Q}}-\bq^{\NC}.\mathbf{\hat{P}})}e^{\frac{i}{\hbar}(\bp^{\prime\;\NC}.\mathbf{\hat{Q}}-\bq^{\prime\;\NC}.\mathbf{\hat{P}})}\eta\rangle\nn\\
&&=e^{-\frac{i}{2\hbar}(\bp^{\NC}.\bq^{\prime\;\NC}-\bp^{\prime\;\NC}.\bq^{\NC})}\langle\eta\vert e^{\left[-\frac{i}{\hbar}(\bp^{\NC}-\bp^{\prime\;\NC}).\mathbf{\hat{Q}}+\frac{i}{\hbar}(\bq^{\NC}-\bq^{\prime\;\NC}).\mathbf{\hat{P}}\right]}\eta\rangle\nn\\
&&=e^{-\frac{i}{2\hbar}(\bp^{\NC}.\bq^{\prime\;\NC}-\bp^{\prime\;\NC}.\bq^{\NC})}\int_{\mathbb{R}^{2}}\overline{\eta(\br)} e^{\left[-\frac{i}{\hbar}(\bp^{\NC}-\bp^{\prime\;\NC}).\mathbf{\hat{Q}}+\frac{i}{\hbar}(\bq^{\NC}-\bq^{\prime\;\NC}).\mathbf{\hat{P}}\right]}\eta(\br)d\br\nn\\
&&=e^{-\frac{i}{2\hbar}\left[(\bp^{\NC}.\bq^{\prime\;\NC}-\bp^{\prime\;\NC}.\bq^{\NC})+(\bp^{\NC}-\bp^{\prime\;\NC}).(\bq^{\NC}-\bq^{\prime\;\NC})\right]}\int_{\mathbb{R}^{2}}\overline{\eta(\br)} e^{-\frac{i}{\hbar}(\bp^{\NC}-\bp^{\prime\;\NC}).\mathbf{\hat{Q}}}\widetilde{\eta}(\br)d\br,\nn\\
\ena
where
\be\label{def-tilde-eta}
\widetilde{\eta}=e^{\frac{i}{\hbar}(\bq^{\NC}-\bq^{\prime\;\NC}).\mathbf{\hat{P}}}\eta.
\en
Therefore, (\ref{rep-ker-proof}) now reads
\bea\label{rep-ker-proof-contd}
\lefteqn{\langle\neta\vert\netap\rangle}\nn\\
&&=e^{-\frac{i}{2\hbar}\left[(\bp^{\NC}.\bq^{\prime\;\NC}-\bp^{\prime\;\NC}.\bq^{\NC})+(\bp^{\NC}-\bp^{\prime\;\NC}).(\bq^{\NC}-\bq^{\prime\;\NC})\right]}\nn\\
&&\times \int_{\mathbb{R}^{2}}\overline{\eta(\br)}e^{-\frac{i}{\hbar}(\bp^{\NC}-\bp^{\prime\NC}).\br}e^{\frac{i}{\hbar}(\bq^{\NC}-\bq^{\prime\;\NC}).\mathbf{\hat{P}}}\eta(\br)d\br\nn\\
&&=e^{-\frac{i}{2\hbar}\left[(\bp^{\NC}.\bq^{\prime\;\NC}-\bp^{\prime\;\NC}.\bq^{\NC})+(\bp^{\NC}-\bp^{\prime\;\NC}).(\bq^{\NC}-\bq^{\prime\;\NC})\right]}\nn\\
&&\times \int_{\mathbb{R}^{2}}\overline{\eta(\br)}e^{-\frac{i}{\hbar}(\bp^{\NC}-\bp^{\prime\NC}).\br}\eta(\br+\bq^{\NC}-\bq^{\prime\;\NC})d\br\nn\\
&&=e^{-\frac{i}{2\hbar}\left[(\bp^{\NC}.\bq^{\prime\;\NC}-\bp^{\prime\;\NC}.\bq^{\NC})+(\bp^{\NC}-\bp^{\prime\;\NC}).(\bq^{\NC}-\bq^{\prime\;\NC})\right]}\nn\\
&&\times\frac{1}{\pi s^{2}}\int_{\mathbb{R}^{2}}e^{-\frac{1}{2s^{2}}\vert\br\vert^{2}}e^{-\frac{i}{\hbar}(\bp^{\NC}-\bp^{\prime\NC}).\br}e^{-\frac{1}{2s^{2}}\vert\br+\bq^{\NC}-\bq^{\prime\;\NC}\vert^{2}}d\br\nn\\
&&=\frac{1}{\pi s^{2}}e^{-\frac{i}{2\hbar}\left[(\bp^{\NC}.\bq^{\prime\;\NC}-\bp^{\prime\;\NC}.\bq^{\NC})+(\bp^{\NC}-\bp^{\prime\;\NC}).(\bq^{\NC}-\bq^{\prime\;\NC})\right]}\nn\\
&&\times\int_{\mathbb{R}^{2}}e^{-\frac{i}{\hbar}(\bp^{\NC}-\bp^{\prime\NC}).\br}e^{-\frac{1}{s^{2}}\vert\br\vert^{2}-\frac{1}{s^{2}}\br.(\bq^{\NC}-\bq^{\prime\;\NC})-\frac{\vert\bq^{\NC}-\bq^{\prime\;\NC}\vert^{2}}{2s^{2}}}d\br\nn\\
&&=\frac{1}{\pi s^{2}}e^{-\frac{i}{2\hbar}\left[(\bp^{\NC}.\bq^{\prime\;\NC}-\bp^{\prime\;\NC}.\bq^{\NC})+(\bp^{\NC}-\bp^{\prime\;\NC}).(\bq^{\NC}-\bq^{\prime\;\NC})\right]}\nn\\
&&\times e^{\left[-\frac{s^{2}}{4\hbar^{2}}\vert\bp^{\NC}-\bp^{\prime\;\NC}\vert^{2}-\frac{1}{4s^{2}}\vert\bq^{\NC}-\bq^{\prime\;\NC}\vert^{2}+\frac{i}{2\hbar}(\bp^{\NC}-\bp^{\prime\;\NC}).(\bq^{\NC}-\bq^{\prime\;\NC)}\right]}\nn\\
&&\times\int_{\mathbb{R}^{2}}e^{-\frac{1}{s^{2}}\left\vert\br+i\frac{s^{2}}{2\hbar}(\bp^{\NC}-\bp^{\prime\;\NC})+\frac{1}{2}(\bq^{\NC}-\bq^{\prime\;\NC})\right\vert^{2}}d\br\nn\\
&&=e^{\left[-\frac{i}{2\hbar}\left(\bp^{\NC}.\bq^{\prime\;\NC}-\bp^{\prime\;\NC}.\bq^{\NC}\right)-\frac{s^{2}}{4\hbar^{2}}\vert\bp^{\NC}-\bp^{\prime\;\NC}\vert^{2}-\frac{1}{4s^{2}}\vert\bq^{\NC}-\bq^{\prime\;\NC}\vert^{2}\right]}\nn\\
\ena
Now writing $\bq^{\NC}$,$\bp^{\NC}$ in terms of $\bq$ and $\bp$ with the help of (\ref{nc-chng-vrbl}) and subsequently using (\ref{local-expnts}), one finally obtains
\bea\label{rep-krnl-fin-exprssn}
\lefteqn{\langle\neta\vert\netap\rangle}\nn\\
&&=e^{\frac{i}{\hbar}\xi((\bq,\bp),(\bq^{\prime},\bp^{\prime}))-\frac{s^{2}}{4\hbar^{2}}\vert\bp-\bp^{\prime}\vert^{2}-\frac{1}{4s^{2}}\vert\bq-\bq^{\prime}\vert^{2}}e^{\frac{i\vartheta}{\hbar^{2}}\xi^{\prime}((\bq,\bp),(\bq^{\prime},\bp^{\prime}))}\nn\\
&&\times e^{-\frac{1}{4s^{2}}\left[\frac{\vartheta^{2}}{\hbar^{2}}(p_{1}-p_{1}^{\prime})^{2}+2\frac{\vartheta}{\hbar}(q_{2}-q_{2}^{\prime})(p_{1}-p_{1}^{\prime})\right]}e^{\frac{i\mathcal{B}}{\hbar^{2}}\xi^{\prime\prime}((\bq,\bp),(\bq^{\prime},\bp^{\prime}))-\frac{s^{2}}{4\hbar^{2}}\left[\frac{\mathcal{B}^{2}}{\hbar^{2}}(q_{2}-q_{2}^{\prime})^{2}+2\frac{\mathcal{B}}{\hbar}(p_{1}-p_{1}^{\prime})(q_{2}-q_{2}^{\prime})\right]}.\nn
\ena
\qed
\begin{remark}\label{rep-krnl-nc-qm-limits}
It is worth remarking here that as $\mathcal{B},\vartheta\rightarrow 0$,
\be\label{rep-krnl-lim-eqn}
K^{\NC}((\bq,\bp),(\bq^{\prime},\bp^{\prime}))\rightarrow e^{\frac{i}{\hbar}\xi((\bq,\bp),(\bq^{\prime},\bp^{\prime}))-\frac{s^{2}}{4\hbar^{2}}\vert\bp-\bp^{\prime}\vert^{2}-\frac{1}{4s^{2}}\vert\bq-\bq^{\prime}\vert^{2}},
\en
which is the canonical reproducing kernel that one obtains for the Weyl-Heisenberg group in 2-dimensions.
\end{remark}

\section{Noncommutative coherent states in the degenerate case}
%and the associated reproducing kernel for the unitary irreducible representations of $\g$ due to $\rho^{2}\alpha^{2}-\gamma\beta\sigma\tau=0$
\label{sec:degenrt-rep}
In the previous section, we have computed the reproducing kernel $K^{\NC}((\bq,\bp),(\bq^{\prime},\bp^{\prime}))$ from the generic unitary irreducible representations of $\g$ due to nonzero $\rho$, $\sigma$ and $\tau$ satisfying  $\rho^{2}\alpha^{2}-\gamma\beta\sigma\tau\neq0$. In this section, using similar arguments, we shall compute the reproducing kernel $\tilde{K}^{\NC}((\bq,\bp),(\bq^{\prime},\bp^{\prime}))$ from the unitary irreducible representations of $\g$ for which each of $\rho$, $\sigma$ and $\tau$ is nonzero and $\rho^{2}\alpha^{2}-\gamma\beta\sigma\tau=0$ holds. This family of representations has been computed in \cite{ncqmjpa}:
\begin{eqnarray}\label{eq:degenerate-rep-momentumspace}
    \lefteqn{(\hat{U}^{\kappa,\delta}_{\rho,\zeta}(\theta,\phi,\psi,q_{1},q_{2},p_{1},p_{2})\hat{f})(s)}\nonumber\\
    &&=e^{i\kappa q_{1}+i\delta q_{2}+i\rho\left(\theta-\alpha q_{1}s-\frac{\alpha q_{1}p_{1}}{2}-\frac{\zeta\alpha^{2}q_{1}q_{2}}{2\beta}\right)}e^{i\frac{\rho}{\zeta}\left(\phi+\beta p_{2}s+\frac{\zeta\alpha q_{2}p_{2}}{2}+\frac{\beta}{2}p_{1}p_{2}\right)}\nonumber\\
    &&\times e^{i\frac{\zeta\rho\alpha^{2}}{\gamma\beta}\psi}\hat{f}(s+p_{1}+\frac{\zeta\alpha q_{2}}{\beta}),
\end{eqnarray}
where $f\in L^{2}(\hat{\mathbb{R}},ds)$. It is to be noted that given  $\rho\neq 0$, an ordered pair $(\kappa,\delta)$ and $\zeta\in (-\infty, 0)\cup(0,\infty)$ satisfying $\rho=\sigma\zeta=\frac{\gamma\beta\tau}{\zeta\alpha^{2}}$, one precisely obtains a unitary irreducible representation of $\g$.

Inverse Fourier transform of (\ref{eq:degenerate-rep-momentumspace}) leads to
\begin{eqnarray}\label{eq:degenerate-rep-momentumspace-inv-trnsfrm}
    \lefteqn{(U^{\kappa,\delta}_{\rho,\zeta}(\theta,\phi,\psi,q_{1},q_{2},p_{1},p_{2})f)(r)}\nonumber\\
    &&=e^{i\rho\left(\theta+\frac{1}{\zeta}\phi+\frac{\zeta\alpha^{2}}{\gamma\beta}\psi\right)+i\kappa q_{1}+i\delta q_{2}-i\rho\alpha rp_{1}-\frac{i\rho\alpha^{2}\zeta}{\beta}rq_{2}+\frac{i\rho\alpha}{2}(q_{1}p_{1}-q_{2}p_{2})}\nonumber\\
    &&\times e^{i\rho\left(\frac{\alpha^{2}\zeta}{2\beta}q_{1}q_{2}-\frac{\beta}{2\zeta}p_{1}p_{2}\right)}f(r-q_{1}+\frac{\beta}{\alpha\zeta}p_{2}),
\end{eqnarray}
where $f\in L^{2}(\mathbb{R},dr)$.

In analogy with section \ref{sec:cs ncqm}, given a fixed vector $\tilde{\chi}\in L^{2}(\mathbb{R},dr)$ and a fixed point $(\bq,\bp)\in\mathbb{R}^{4}$, let us first define the vector in $L^{2}(\mathbb{R},dr)$ as:
\be\label{def-nc-cs-deg}
\nnchi=U^{\kappa,\delta}_{\rho,\zeta}(0,0,0,-\bq,\bp)\chi.
\en
With the following change of variables:
\beno
\hbar=\frac{1}{\rho\alpha},\;\;\vartheta=-\frac{\sigma\beta}{\rho^{2}\alpha^{2}},
\enno
and recalling that $\rho=\sigma\zeta=\frac{\gamma\beta\tau}{\zeta\alpha^{2}}$ holds, one can rewrite $\nnchi$ suitably as:
\be
\nnchi(r)=e^{-i\kappa q_{1}-i\delta q_{2}-\frac{ir}{\hbar}\left(p_{1}+\frac{\hbar}{\vartheta}q_{2}\right)-\frac{i}{2\hbar}(q_{1}p_{1}-q_{2}p_{2})+\frac{i}{2}\left(-\frac{q_{1}q_{2}}{\vartheta}+\frac{\vartheta}{\hbar^{2}}p_{1}p_{2}\right)}\tilde{\chi}\left(r+q_{1}-\frac{\vartheta}{\hbar}p_{2}\right).
\en
\bedefin\label{cs-degnrt-def}
Define the degenerate noncommutative coherent states as the following vectors in $L^{2}(\mathbb{R},dr)$:
\be\label{eq:normlzd-deg-cohrnt-states}
\nneta(r)=e^{-i\kappa q_{1}-i\delta q_{2}-\frac{ir}{\hbar}\left(p_{1}+\frac{\hbar}{\vartheta}q_{2}\right)-\frac{i}{2\hbar}(q_{1}p_{1}-q_{2}p_{2})+\frac{i}{2}\left(-\frac{q_{1}q_{2}}{\vartheta}+\frac{\vartheta}{\hbar^{2}}p_{1}p_{2}\right)}\tilde{\eta}\left(r+q_{1}-\frac{\vartheta}{\hbar}p_{2}\right),
\en
where $\tilde{\eta}$ is a vector given by $\tilde{\eta}=\frac{\tilde{\chi}}{\vert\vert\tilde{\chi}\vert\vert}$.
\findefi

Now the underlying observables are functions in the Hilbert space
$L^{2}(\mathbb{R}^{4},d\tilde\nu(\bq,\bp))$ equipped with the measure
\begin{equation} \label{tilde-nu}
 d\tilde\nu(\bq,\bp)=\frac{1}{2\pi\hbar}dq_{1}dp_{1}d\mu(q_{2},p_{2}),
\end{equation}
where $d\mu$ is an arbitrary probability measure on~$\mathbb R^2$.

\belem\label{lem-res-idnty-deg-cohrnt-states}
The vectors $\nneta$ defined as the degenerate noncommutative coherent states by (\ref{eq:normlzd-deg-cohrnt-states}) satisfy the following integral relation:
\be\label{stmnt-res-idnty-deg}
\int_{\mathbb{R}^{4}}\vert\nneta\rangle\langle\nneta\vert\;d\tilde{\nu}(\bq,\bp)=\mathbb{I},
\en
where $\mathbb{I}$ is the identity operator on $L^{2}(\mathbb{R},dr)$.
\enlem
\prf{.}
Using the following change of variables:
\be\label{nc-chng-vrbl-dgnrt-case}
\begin{aligned}
\nQ&=q_{1}-\frac{\vartheta}{\hbar}p_{2}\\
\nR&=q_{2}\\
\nS&=p_{1}+\frac{\hbar}{\vartheta}q_{2}+2\kappa\hbar\\
\nT&=p_{2},
\end{aligned}
\en
one can rewrite the degenerate noncommutative coherent states (\ref{eq:normlzd-deg-cohrnt-states}) as:
\be\label{nc-cohrnt-states-deg-nc-coordnts}
\nneta(r)=e^{-i\delta q_{2}^{\NC}-\frac{ir}{\hbar}(p_{1}^{\NC}-2\kappa\hbar)-\frac{i}{2\hbar}q_{1}^{\NC}p_{1}^{\NC}-\frac{i\vartheta}{\hbar}\kappa p_{2}^{\NC}}\tilde{\eta}(r+q_{1}^{\NC}).
\en
Observe that $dq_1 dp_1=dq_1^{\NC}dp_1^{\NC}$,
by~(\ref{nc-chng-vrbl-dgnrt-case}).
Thus for any $f,g\in L^{2}(\mathbb{R},dr)$, we obtain
\bea
\lefteqn{\int_{\mathbb{R}^{4}}\langle f\vert\nneta\rangle\langle\nneta\vert g\rangle\;d\tilde{\nu}(\bq,\bp)}\nn\\
&&=\frac{1}{2\pi\hbar}\int_{\mathbb{R}^{4}}\left[\int_{\mathbb{R}}\int_{\mathbb{R}}\overline{f(r)}e^{-\frac{i}{\hbar}(r-r^{\prime})(\nS-2\kappa\hbar)}\tilde{\eta}(r+\nQ)\overline{\tilde{\eta}(r^{\prime}+\nQ)}g(r^{\prime})dr dr^{\prime}\right]\nn\\
&&\qquad\qquad\times d\nQ d\nS d\mu(\nR,\nT)\nn\\
&&=\int_{\mathbb{R}^{3}}\left[\int_{\mathbb{R}}\int_{\mathbb{R}}\delta(r-r^{\prime})\overline{f(r)}\tilde{\eta}(r+\nQ)\overline{\tilde{\eta}(r^{\prime}+\nQ)}g(r^{\prime})dr\;dr^{\prime}\right] d\nQ d\mu(\nR,\nT)\nn\\
&&=\vert\vert\tilde{\eta}\vert\vert^{2}\int_{\mathbb{R}^{2}}\left[\int_{\mathbb{R}}\overline{f(r)}g(r)dr\right] d\mu(\nR,\nT)\nn\\
&&=\langle f\vert g\rangle .
\ena
\qed

The reproducing kernel associated with the degenerate noncommutative coherent states will be given by 
\be\label{eq:deg-cohrnt-states-rep-kern}
\tilde{K}^{\NC}((\bq,\bp),(\bq^{\prime},\bp^{\prime}))=\langle\nneta\vert\nnneta\rangle.
\en
This degenerate reproducing kernel is explicitly computed in the following proposition:
\begin{Prop}
Provided one chooses the ground state vector $\tilde{\eta}\in L^{2}(\mathbb{R},dr)$ in (\ref{eq:normlzd-deg-cohrnt-states}) to be the following normalized Gaussian function:
\be\label{grnd-state-gaussn-deg-rep}
\tilde{\eta}(r)=\frac{1}{{\pi}^{\frac14}s^{\frac12}}e^{-\frac{r^{2}}{2s^{2}}},
\en
then, the reproducing kernel $\tilde{K}^{\NC}$ (see (\ref{eq:deg-cohrnt-states-rep-kern})) associated with the Lie group $\g$ reads
\bea\label{rep-krnl-dfn-grp-ncqm-degnrt-rep}
\lefteqn{\tilde{K}^{\NC}((\bq,\bp),(\bq^{\prime},\bp^{\prime}))}\nn\\
&&=e^{\frac{i}{\hbar}\xi((\bq,\bp),(\bq^{\prime},\bp^{\prime}))+\frac{i}{\vartheta}\xi^{\prime\prime}((\bq,\bp),(\bq^{\prime},\bp^{\prime}))+\frac{i\vartheta}{\hbar^{2}}\xi^{\prime}((\bq,\bp),(\bq^{\prime},\bp^{\prime}))+i\kappa(q_{1}-q_{1}^{\prime})+i\delta(q_{2}-q_{2}^{\prime})-\frac{1}{4s^{2}}(q_{1}-q_{1}^{\prime})^{2}}\nn\\
&&\times e^{-\frac{s^{2}}{4\vartheta^{2}}(q_{2}-q_{2}^{\prime})^{2}-\frac{s^{2}}{4\hbar^{2}}(p_{1}-p_{1}^{\prime})^{2}-\frac{\vartheta^{2}}{4s^{2}\hbar^{2}}(p_{2}-p_{2}^{\prime})^{2}+\frac{\vartheta}{2s^{2}\hbar}(q_{1}p_{2}-q_{1}p_{2}^{\prime}-q_{1}^{\prime}p_{2}+q_{1}^{\prime}p_{2}^{\prime})}\nn\\
&&\times e^{-\frac{s^{2}}{2\hbar\vartheta}(p_{1}q_{2}-p_{1}q_{2}^{\prime}-p_{1}^{\prime}q_{2}+p_{1}^{\prime}q_{2}^{\prime})},
\ena
where $\xi$, $\xi^{\prime}$ and $\xi^{\prime\prime}$ are all given by (\ref{local-expnts}). Also, $s$ stands for the standard deviation associated with the position coordinate $r$ and hence has the dimension of length.
\end{Prop}

% The proof of the above proposition requires one to write down $\nneta(r)$
% explicitly using (\ref{eq:normlzd-deg-cohrnt-states}) and
% (\ref{grnd-state-gaussn-deg-rep}) and finally substituting the resulting
% expression in (\ref{eq:deg-cohrnt-states-rep-kern}).
% It is rather straightforward and hence is omitted.

\prf.
Using (\ref{nc-cohrnt-states-deg-nc-coordnts}) we get
\begin{align*}
\langle\nneta|\nnneta\rangle
&= e^{-i\delta(q_2^{\prime\NC}-q_2^{\NC})
 -\frac{i}{2\hbar}
   (q_1^{\prime\NC}p_1^{\prime\NC}-q_1^{\NC}p_1^{\NC})
 -\frac{i\vartheta}{\hbar}\kappa
   (p_2^{\prime\NC}-p_2^{\NC})}   \\
&\qquad\times
 \int_{\mathbb R}
   e^{-\frac{ir}{\hbar}(p_1^{\prime\NC}-p_1^{\NC})}
   \tilde{\eta}(r+q_1^{\prime\NC})
   \overline{\tilde{\eta}(r+q_1^{\NC})} \, dr   \\
&= e^{-i\delta(q_2^{\prime\NC}-q_2^{\NC})
 -\frac{i}{2\hbar}
   (q_1^{\prime\NC}p_1^{\prime\NC}-q_1^{\NC}p_1^{\NC})
 -\frac{i\vartheta}{\hbar}\kappa
   (p_2^{\prime\NC}-p_2^{\NC})}   \\
&\qquad\times
 \frac1{\pi^{1/2}s} \int_{\mathbb R}
   e^{-\frac{ir}{\hbar}(p_1^{\prime\NC}-p_1^{\NC})}
   e^{-\frac{(r+q_1^{\prime\NC})^2}{2s^2}
      -\frac{(r+q_1^{\NC})^2}{2s^2}} \, dr   \\
&= e^{-i\delta(q_2^{\prime\NC}-q_2^{\NC})
 -\frac{i}{2\hbar}
   (q_1^{\prime\NC}p_1^{\prime\NC}-q_1^{\NC}p_1^{\NC})
 -\frac{i\vartheta}{\hbar}\kappa
   (p_2^{\prime\NC}-p_2^{\NC})}   \\
&\qquad\times
    e^{-\frac{\hbar^2(q_1^{\NC}-q_1^{\prime\NC})^2
              +2i\hbar s^2 (p_1^{\NC}-p_1^{\prime\NC})
                      (q_1^{\NC}+q_1^{\prime\NC})
              + s^4 (p_1^{\NC}-p_1^{\prime\NC})^2 } {4h^2s^2}} ,
\end{align*}
by~the familiar formula for Gaussian integrals
$$ \int_{\mathbb R} e^{ar-b-cr^2}\,dr =
 \pi^{1/2} c^{-1/2} e^{\frac{a^2}{4c}-b} ,  \qquad c>0 .    $$
A~routine manipulation gives~(\ref{rep-krnl-dfn-grp-ncqm-degnrt-rep}).
\qed

\section{Toeplitz operators and semiclassical limits}\label{secTOEP}
We~now proceed to consider a variant of the well-known Berezin-Toeplitz
quantization procedure in the context of the coherent states,
and resolution of the identity, from the preceding section.
Our~strategy will be to relate the corresponding Toeplitz operators
(defined in (\ref{defTOEP}) below) to~the analogous operators in
the standard setting.

Consider the Fock space
$$ \FFh:=\{f\in L^2(\CC^2,e^{-|\bz|^2}(\pi\hbar)^{-2}\,dA(\bz)):
f\text{ is holomorphic on }\CC^2\}    $$
(here~$dA$ stands for the Lebesgue area measure), and let $V$ be the map
$$ Vf(\bq,\bp) := e^{-|\bznc|^2/{2\hbar}} f(\bznc),   \qquad f\in\FFh,   $$ 
where we have introduced the notation
\begin{equation}
 \bznc_j = \sqrt{\frac\hbar2} \frac{\bq_j^{\NC}}s
 - \frac{is\bp_j^{\NC}}{\sqrt{2\hbar}} , \qquad j=1,2.   \label{XXX}
\end{equation}
From the equality
\begin{equation} \label{MEA}
 dA(\bznc) = \frac{d\bp^{\NC} d\bq^{\NC}}4
 = \frac{|\hbar^2-\cB\vthta|}{4\hbar^2} \,d\bp\,d\bq
 = \pi^2 \hbar^2 \,d\nu(\bq,\bp),
\end{equation}
where as before
\begin{equation} \label{MEB}
 d\nu(\bq,\bp) = \frac{|\hbar^2-\cB\vthta|}{4\pi^2\hbar^4}\,d\bq\,d\bp,  
\end{equation}
one~verifies that $V$~is an isometry from $\FFh$ into $L^2(\RR^4,d\nu(\bq,\bp))$.
If~$\{e_n\}_n$ is an arbitrary orthonormal basis of~$\FFh$, then $\{Ve_n\}_n$
will be an orthonormal basis of the image $\operatorname{Ran}V=:\WW'$ of~$V$.
Using the standard formula for a reproducing kernel in terms of an orthonormal
basis~\cite{Arns}, and the fact that the reproducing kernel of $\FFh$ is well
known to be given by $K_{\FFh}(\mathbf x,\mathbf y)=
e^{\spr{\mathbf x,\mathbf y}/\hbar}$, we~see that $\WW'$ is a reproducing
kernel Hilbert space (RKHS) with reproducing kernel 
\begin{align*}
K_{\WW'}((\bq,\bp),(\bq',\bp')) &= \sum_n Ve_n(\bznc) \overline{Ve_n(\bZnc)} \\
&= e^{-|\bznc|^2/2\hbar-|\bZnc|^2/2\hbar} \sum_n e_n(\bznc) \overline{e_n(\bZnc)}\\
&= e^{-|\bznc|^2/2\hbar-|\bZnc|^2/2\hbar} e^{\spr{\bznc,\bZnc}/\hbar}   \\
&= K^{\NC}((\bq,\bp),(\bq',\bp')), 
\end{align*}
upon a short computation (cf.~(\ref{rep-ker-proof-contd})).
Since a RKHS is uniquely determined by its reproducing kernel~\cite{Arns},
it~follows that in fact $\WW'=\WW$.
Thus $V$ is a unitary isomorphism of $\FFh$ onto~$\WW$.

Recall that for $f\in L^\infty(\CC^2)$, the Toeplitz operator $T_f$ on $\FFh$
is given~by
$$ T_f u = P(fu), \qquad u\in\FFh ,    $$
where $P:L^2(e^{-|\bz|^2/\hbar}(\pi\hbar)^{-2}\,dA(\bz))\to\FFh$ is the orthogonal
projection. Alternatively, $T_f$~is determined by the property that
$$ \spr{T_f u,v} = \int_{\CC^2} f(\bz) u(\bz) \overline{v(\bz)}
 e^{-|\bz|^2/\hbar} \, \frac{d\bz}{(\pi\hbar)^2}  \qquad \forall u,v\in\FFh. $$
Similarly, we~have Toeplitz operators~$\cT_F$, $F\in L^\infty(\RR^4)$,
on~$\WW$ defined~by 
\be
 \cT_F u = \cP(Fu),  \qquad u\in\WW ,    \label{defTOEP}
\en
where $\cP:L^2(\RR^4,d\nu)\to\WW$ is the orthogonal projection;
alternatively, $\cT_F$~is determined by the property that
$$ \spr{\cT_F u,v} = \int_{\RR^4} F(\bq,\bp) u(\bq,\bp) \overline{v(\bq,\bp)}
 \,d\nu(\bq,\bp)  \qquad \forall u,v\in\WW .   $$
Now~by a simple change of variable (cf.~(\ref{MEA}))
\begin{align*}
\spr{\cT_F Vu,Vv}
&= \int_{\RR^4} F(\bq,\bp) Vu(\bq,\bp) \overline{Vv(\bq,\bp)} \,d\nu(\bq,\bp)\\
&= \int_{\RR^4} F(\bq,\bp) e^{-|\bznc|^2/\hbar} u(\bznc) \overline{v(\bznc)}
 \,d\nu(\bq,\bp)   \\
&= \int_{\CC^2} F(\iota(\bznc)) e^{-|\bznc|^2/\hbar} u(\bznc) \overline{v(\bznc)}
 \,\frac{dA(\bznc)}{(\pi\hbar)^2},
\end{align*}
for~all $u,v\in\FFh$, where $\iota$ is the inverse to the coordinate
transformation (\ref{XXX}): 
\begin{equation} \label{IOTA}
\begin{gathered}
 \iota(\bznc) := \Big(q_1^{\NC},
 \frac{\hbar^2 q_2^{\NC}-\hbar\vthta p_1^{\NC}}{\hbar^2-\cB\vthta},
 \frac{\hbar^2 p_1^{\NC}-\hbar\cB q_2^{\NC}}{\hbar^2-\cB\vthta},
 p_2^{\NC}\Big) \in\RR^4,  \\
 \bq^{\NC}=\frac s{\sqrt{2\hbar}}(\bznc+\overline{\bznc}), \quad
 \bp^{\NC}=\frac{\sqrt{\hbar/2}}{is}(\overline{\bznc}-\bznc) . 
\end{gathered}
\end{equation}
Consequently,
\begin{equation}
 V^* \cT_F V = T_{F\circ\iota}.   \label{VTV}
\end{equation}

\begin{remark}
From the last formula one can see what are the commutators of the Toeplitz
operators $\cT_{p_j},\cT_{q_k}$, $j=1,2$, on~$L^2(\RR^4,d\nu)$.
Indeed, from the formulas for the Toeplitz operators on~$\FFh$,
$$ T_{z_j}=z_j, \qquad T_{\overline z_j}=h\frac\partial{\partial z_j},  $$
and the resulting commutator identity
$$ [T_{z_j},T_{\overline z_k}] = -\delta_{jk}\hbar \id,   $$
one gets using (\ref{VTV})
\begin{align*}
[\cT_{p_1},\cT_{p_2}] &= -\frac{i\cB\hbar^2}{\hbar^2-\cB\vthta}\id, \\
[\cT_{p_1},\cT_{q_1}] &= [\cT_{p_2},\cT_{q_2}] = 
 -\frac{i\hbar^3}{\hbar^2-\cB\vthta}\id, \\
[\cT_{q_1},\cT_{q_2}] &= -\frac{i\hbar^2\vthta}{\hbar^2-\cB\vthta}\id ,
\end{align*}
all remaining commutators being zero.
Note that these differ from the commutator relations for the corresponding
quantum observables (cf.~eqn.~(3.7) in~\cite{ncqmjpa})
$$ [\hat Q_j,\hat P_k]=i\hbar\delta_{jk}\id, \quad
[\hat Q_1,\hat Q_2]=i\vthta\id, \quad [\hat P_1,\hat P_2]=i\cB\id  $$
by a factor of $\frac{\hbar^2}{\cB\vthta-\hbar^2}$.
\end{remark}

Now~from the Berezin-Toeplitz quantization (see~e.g.~\cite{Schl},~\cite{AE}),
it~is known that for $f,g$, say, smooth with compact support, 
one~has the asymptotic expansion
$$ T_f T_g \approx \sum_{j=0}^\infty \hbar^j T_{C_j(f,g)}    $$
as $\hbar\searrow0$, in~the sense of operator norms, where
$$ C_j(f,g) = (-1)^j \sum_{|\alpha|=j} \frac1{\alpha!}
 \frac{\partial^\alpha f}{\partial z^\alpha}
 \frac{\partial^\alpha g}{\partial \overline z^\alpha}   $$
(here the summation is over all multiindices $\alpha\in\mathbf N^2$
of length~$j$). From the computation
\begin{align*}
V^*\cT_F\cT_G V &= (V^*\cT_F V)(V^*\cT_G V) = T_{F\circ\iota}T_{G\circ\iota}\\
&\approx \sum_{j=0}^\infty \hbar^j T_{C_j(F\circ\iota,G\circ\iota)}
 = \sum_{j=0}^\infty \hbar^j
 V^* \cT_{\cC_j(F\circ\iota,G\circ\iota)\circ\iota^{-1}}V ,
\end{align*}
we~thus see that we have an asymptotic expansion,
in~the sense of operator norms, 
$$ \cT_F\cT_G \approx \sum_{j=0}^\infty \hbar^j \cT_{\cC_j(F,G)}   $$
as $\hbar\searrow0$, with 
\begin{equation}
 \cC_j(F,G) := C_j(F\circ\iota,G\circ\iota)\circ\iota^{-1} .  \label{CCJ}
\end{equation}
This gives rise also to the associated star-product
$$ F*G := \sum_{j=0}^\infty \hbar^j \cC_j(F,G) ,   $$
so~that, heuristically, $\cT_F\cT_G\approx\cT_{F*G}$.

In~particular, $\cC_0(F,G)=FG$ (the~pointwise product), while
$$ \cC_1(F,G) = 
 \begin{bmatrix}
  \dfrac{\partial F}{\partial p_1},
  \dfrac{\partial F}{\partial p_2},
  \dfrac{\partial F}{\partial q_1},
  \dfrac{\partial F}{\partial q_2}  \end{bmatrix}
  \cdot A \cdot
 \begin{bmatrix} \dfrac{\partial G}{\partial p_1} \\
  \dfrac{\partial G}{\partial p_2} \\
  \dfrac{\partial G}{\partial q_1} \\
  \dfrac{\partial G}{\partial q_2} \end{bmatrix} ,   $$
where
$$ A = \begin{bmatrix}
 -\dfrac{\hbar(\hbar^4+\cB^2 s^4)}{2s^2(\hbar^2-\cB\vthta)^2}  &
 \dfrac{-i\cB \hbar}{2(\hbar^2-\cB\vthta)} &
 \dfrac{-i\hbar^2}  {2(\hbar^2-\cB\vthta)} &
 \dfrac{\hbar^2(\cB s^4+\hbar^2\vthta)}{2s^2(\hbar^2-\cB\vthta)^2}  \\
 \dfrac{i\cB\hbar}{2(\hbar^2-\cB\vthta)}  &
 -\dfrac \hbar{2s^2}  & 0 &
 \dfrac{-i\hbar^2}{2(\hbar^2-\cB\vthta)}  \\
 \dfrac{i\hbar^2}{2(\hbar^2-\cB\vthta)} &
 0 & -\dfrac{s^2}{2\hbar}  &
 \dfrac{-i\hbar\vthta}{2(\hbar^2-\cB\vthta)}  \\
 \dfrac{\hbar^2(\cB s^4+\hbar^2\vthta)}{2s^2(\hbar^2-\cB\vthta)^2}  &
 \dfrac{i\hbar^2}{2(\hbar^2-\cB\vthta)} & 
 \dfrac{i\hbar\vthta}{2(\hbar^2-\cB\vthta)} &
 -\dfrac{\hbar^3(s^4+\vthta^2)}{2s^2(\hbar^2-\cB\vthta)^2}
\end{bmatrix} .   $$
The~last matrix looks much nicer in terms of the ``renormalized'' parameters
$$ B=\frac{\cB}\hbar, \quad T:=\frac\vthta\hbar,
 \quad S:=\frac s{\sqrt{\hbar}} ;   $$
namely,
$$ A = \begin{bmatrix}
 -\dfrac{1+B^2S^4}{2S^2(1-BT)^2} & -\dfrac{iB}{2(1-BT)} &
  -\dfrac i{2(1-BT)} & \dfrac{BS^4+T}{2S^2(1-BT)^2} \\
 \dfrac{iB}{2(1-BT)} & -\dfrac1{2S^2} & 0 & -\dfrac i{2(1-BT)} \\
 \dfrac i{2(1-BT)} & 0 & -\dfrac{S^2}2 & -\dfrac{iT}{2(1-BT)}  \\
 \dfrac{BS^4+T}{2S^2(1-BT)^2} & \dfrac i{2(1-BT)} & \dfrac{iT}{2(1-BT)} &
  -\dfrac{S^4+T^2}{2S^2(1-BT)^2}
\end{bmatrix} .   $$
Note that both $A$ and the inverse transform $\iota$ depend only on $B,T$
and~$S$, but not~on~$\hbar$.
The~parameter~$S$, which arises solely from the choice of the vector~$\eta$
in~(\ref{grnd-state-gaussn}), is~in a sense responsible only for re-scaling
the Planck constant~$\hbar$, and we can choose $S=1$.
\footnote{Note that as far as length only is concerned, $s$~has the same
physical dimension as~$\sqrt{\hbar}$.}
It~then follows from (\ref{CCJ}) that all $\cC_j$, $j\ge0$, will be
bidifferential operators with coefficients given by expressions involving
only $B,T$, in~fact, by~rational functions in $B,T$ with powers of $1-BT$
as the denominators. Replacing the latter by their Taylor expansions around
$(B,T)=(0,0)$, we~thus obtain a joint asymptotic expansion for the product
$F*G$ as $(\hbar,B,T)\to(0,0,0)$. Its~beginning looks as follows
\begin{align*}
F*G &\approx FG - \frac\hbar2 \sum_{k=1}^2 
  \Big(\frac{\partial F}{\partial p_k}-i\frac{\partial F}{\partial q_k}\Big)
  \Big(\frac{\partial G}{\partial p_k}+i\frac{\partial G}{\partial q_k}\Big) \\
& \qquad\qquad
 + \frac{B\hbar}2 \Big[ i\frac{\partial G}{\partial p_1}
    \Big(\frac{\partial F}{\partial p_2}-i\frac{\partial F}{\partial q_2}\Big)
        -i \frac{\partial F}{\partial p_1}
    \Big(\frac{\partial G}{\partial p_2}+i\frac{\partial G}{\partial q_2}\Big)
   \Big]  \\
& \qquad\qquad
 + \frac{T\hbar}2 \Big[ \frac{\partial F}{\partial q_2}
    \Big(\frac{\partial G}{\partial p_1}+i\frac{\partial G}{\partial q_1}\Big)
        + \frac{\partial G}{\partial q_2}
    \Big(\frac{\partial F}{\partial p_1}-i\frac{\partial F}{\partial q_1}\Big)
   \Big]  \\
& \qquad\qquad
 + \frac{\hbar^2}8 \Big[ \sum_{k=1}^2
    \Big(\frac\partial{\partial p_k}+i\frac\partial{\partial q_k}\Big)^2 G
    \cdot
    \Big(\frac\partial{\partial p_k}-i\frac\partial{\partial q_k}\Big)^2 F \\
& \hskip1em
    + 2 \Big(\frac\partial{\partial p_1}+i\frac\partial{\partial q_1}\Big)
        \Big(\frac\partial{\partial p_2}+i\frac\partial{\partial q_2}\Big) G
    \cdot
        \Big(\frac\partial{\partial p_1}-i\frac\partial{\partial q_1}\Big)
        \Big(\frac\partial{\partial p_2}-i\frac\partial{\partial q_2}\Big) F
    \Big]  \\
& \qquad\qquad
  + O((h+B+T)^3).
\end{align*}
(Here the differentiations stop at each~$\cdot$,
i.e.~$\partial F\cdot\partial G$ means $(\partial F)(\partial G)$.)

For~the corresponding commutator, we~get
\begin{align*}
F*G-G*F &\approx
i\hbar \sum_{k=1}^2  \Big(
  \frac{\partial F}{\partial q_k} \frac{\partial G}{\partial p_k} -
  \frac{\partial F}{\partial p_k} \frac{\partial G}{\partial q_k}  \Big)  \\
& \qquad\qquad
 + i\hbar B  \Big(
  \frac{\partial F}{\partial p_2} \frac{\partial G}{\partial p_1} - 
  \frac{\partial F}{\partial p_1} \frac{\partial G}{\partial p_2}  \Big)  \\
& \qquad\qquad
 + i\hbar T  \Big(
  \frac{\partial F}{\partial q_2} \frac{\partial G}{\partial q_1} - 
  \frac{\partial F}{\partial q_1} \frac{\partial G}{\partial q_2}  \Big)  \\
& \qquad\qquad
 + \frac{\hbar^2}8  \Big[
     2 (\partial_{+1}\partial_{+2}G) (\partial_{-1}\partial_{-2}F) -
     2 (\partial_{+1}\partial_{+2}F) (\partial_{-1}\partial_{-2}G)   \\
& \hskip6em
     + \sum_{k=1}^2 \Big(
        (\partial_{+k}^2 G)(\partial_{-k}^2 F) -
        (\partial_{+k}^2 F)(\partial_{-k}^2 G)  \Big)  \Big]    \\
& \qquad\qquad
  + O((h+B+T)^3) ,
\end{align*}
where for the sake of brevity, we~have denoted $\partial_{\pm k}=
\frac\partial{\partial p_k}\pm i\frac\partial{\partial q_k}$, $k=1,2$.
The~first term~is $i\hbar\{F,G\}$, the~Poisson bracket of $F$ and~$G$,
which takes care of the correct semi-classical behavior as $\hbar\to0$;
note that it does not contain any $B$ and~$T$, which come only in the
second-order terms.

\section{Toeplitz operators and semiclassical limits: the degenerate case}
\label{secTOEPdeg}

In~a~similar way as in the preceding section, we~can also treat the
``degenerate'' representation and kernel from Section~\ref{sec:degenrt-rep}.
This time, we~need the Fock space just on the complex plane,
$$ \tFFh:=\{f\in L^2(\CC,e^{-|z|^2}(\pi\hbar)^{-1}\,dA(z)):
f\text{ is holomorphic on }\CC\},   $$
and let $\tV$ be the map
$$ \tV f(\bq,\bp) := e^{i\delta q_2+i\kappa\vthta p_2/\hbar}
  e^{-|\znc|^2/{2\hbar}} f(\znc),   \qquad f\in\tFFh,   $$ 
where
\begin{equation}
 \znc = \sqrt{\dfrac\hbar2} \frac{q_1^{\NC}}s 
 - \frac{is p_1^{\NC}}{\sqrt{2\hbar}} ,    \label{defZNC}
\end{equation}
with $p_1^{\NC},q_1^{\NC}$ given by~(\ref{nc-chng-vrbl-dgnrt-case}).
One~verifies that $\tV$~is an isometry from $\tFFh$ into $L^2(\RR^4,d\tnu)$:
\begin{align*}
\int_{\RR^4} |\tV f(\bq,\bp)|^2 \,d\tnu(\bq,\bp)
&= \frac1{2\pi\hbar} \int_{\RR^4} |f(\znc)|^2 e^{-|\znc|^2/\hbar} 
 \, dq_1^{\NC}\,dp_1^{\NC}\,d\mu(q_2,p_2)   \\
&= \frac1{2\pi\hbar} \int_{\RR^2} |f(\znc)|^2 e^{-|\znc|^2/\hbar} 
 \,dq_1^{\NC}\,dp_1^{\NC}   \\
&= \frac1{\pi\hbar} \int_{\CC} |f(\znc)|^2 e^{-|\znc|^2/\hbar} \,dA(\znc),
\end{align*}
since $dA(\znc)=dp_1^{\NC}dq_1^{\NC}/2$.
Thus if~$\{e_n\}_n$ is an arbitrary orthonormal basis of~$\tFFh$,
then $\{\tV e_n\}_n$ will be an orthonormal basis of the image
$\operatorname{Ran}\tV=:\tWW'$ of~$\tV$. 
Using the standard formula for a reproducing kernel in terms of an orthonormal
basis, and the fact that the reproducing kernel of $\tFFh$ is well known to be
given by $K_{\tFFh}(x,y)=e^{x\overline y/\hbar}$,
we~see that $\tWW'$ is a RKHS with reproducing kernel 
\begin{align*}
K_{\tWW'}((\bq,\bp),(\bq',\bp'))
&= \sum_n \tV e_n(\znc) \overline{\tV e_n(\Znc)} \\
&= e^{i\delta(q_2-q'_2)+i\kappa\vthta(p_2-p'_2)/\hbar}
   e^{-|\znc|^2/2\hbar-|\Znc|^2/2\hbar}
   \sum_n e_n(\znc) \overline{e_n(\Znc)}  \\
&= e^{i\delta(q_2-q'_2)+i\kappa\vthta(p_2-p'_2)/\hbar}
   e^{-|\znc|^2/2\hbar-|\Znc|^2/2\hbar} e^{\znc\overline{\Znc}/\hbar}   \\
&= \tilde K^{\NC}((\bq,\bp),(\bq',\bp')), 
\end{align*}
again upon a short computation. As~before, it~follows that $\tWW'=\tWW$,
the~space for which $\tilde K^{\NC}$ is the reproducing kernel.
Thus $\tV$ is a unitary isomorphism of $\tFFh$ onto~$\tWW$.

The Toeplitz operators~$\tcT_F$, $F\in L^\infty(\RR^4,d\tnu)$,
on~$\tWW$ are now related to the Toeplitz operators $T_f$,
$f\in L^\infty(\CC)$, on~$\tFFh$ via
\begin{align*}
\spr{\tcT_F \tV u,\tV v}
&= \int_{\RR^4} F(\bq,\bp) \tV u(\bq,\bp) \overline{\tV v(\bq,\bp)}
 \,d\tnu(\bq,\bp)  \\
&= \frac1{2\pi\hbar} \int_{\RR^4} F(\bq,\bp) e^{-|\znc|^2/\hbar} u(\znc)
 \overline{v(\znc)} \,dq_1^{\NC}\,dp_1^{\NC}\,d\mu(q_2,p_2)   \\
&= \frac1{2\pi\hbar} \int_{\RR^2}
 \Big(\int_{\RR^2}
 F(q_1^{\NC}+\frac\vthta\hbar p_2,q_2,
 p_1^{\NC}-\frac\hbar\vthta q_2-2\kappa\hbar,p_2) \,d\mu(q_2,p_2) \Big)\\
&\qquad\qquad\times 
 u(\znc) \overline{v(\znc)} e^{-|\znc|^2/\hbar} 
  \,dq_1^{\NC}\,dp_1^{\NC} \\
&= \int_{\CC} \varrho F(\znc) u(\znc) \overline{v(\znc)}
 e^{-|\znc|^2/\hbar} \,\frac{dA(\znc)}{\pi\hbar}  \\
&= \spr{T_{\varrho F} u,v} ,
\end{align*}
for~all $u,v\in\tFFh$, where $\varrho$ is the mapping
$$ \begin{gathered}
 \varrho F(\znc) := \int_{\RR^2}
 F(q_1^{\NC}+\frac\vthta\hbar p_2,q_2,
   p_1^{\NC}-\frac\hbar\vthta q_2-2\kappa\hbar,p_2) \,d\mu(q_2,p_2), \\
 q_1^{\NC}= \frac s{\sqrt{2\hbar}} (\znc+\overline\znc), \quad
 p_1^{\NC}= \frac{\sqrt{\hbar/2}}{is} (\overline\znc-\znc). 
\end{gathered}  $$
Consequently,
\begin{equation}
 \tV^* \tcT_F \tV = T_{\varrho F}.    \label{TDEG}
\end{equation}

\begin{remark}
From the last formula one can again get the commutator relations for the
Toeplitz operators $\tcT_{p_j},\tcT_{q_k}$, $j=1,2$, on~$L^2(\RR^4,d\tnu)$.
This time we get $\tcT_{p_2}=(\int p_2\,d\mu(q_2,p_2))\id$,
$\tcT_{q_2}=(\int q_2\,d\mu(q_2,p_2))\id$ and
$$ [\tcT_{q_1},\tcT_{p_1}]=-[\tcT_{p_1},\tcT_{q_1}]=i\hbar\id,   $$
all remaining commutators being zero.
\end{remark}

Using again the asymptotic expansion known from the Berezin-Toeplitz
quantization:
$$ T_f T_g \approx \sum_{j=0}^\infty \hbar^j T_{\tilde C_j(f,g)}    $$
as $\hbar\searrow0$, in~the sense of operator norms, where
$$ \tilde C_j(f,g) = \frac{(-1)^j}{j!}
 \frac{\partial^j f}{\partial z^j}
 \frac{\partial^j g}{\partial \overline z^j} ,   $$
one sees from
$$ \tV^*\tcT_F\tcT_G\tV = T_{\varrho F}T_{\varrho G}
\approx \sum_{j=0}^\infty \hbar^j T_{\tilde C_j(\varrho F,\varrho G)}
 = \sum_{j=0}^\infty \hbar^j
 \tV^* \tcT_{\varrho^* \tilde C_j(\varrho F,\varrho G)}\tV , $$
that there is an asymptotic expansion, in~the sense of operator norms, 
$$ \tcT_F\tcT_G \approx \sum_{j=0}^\infty \hbar^j \tcT_{\tcC_j(F,G)}   $$
as $\hbar\searrow0$, with 
\begin{equation}
 \tcC_j(F,G) := \varrho^* \tilde C_j(\varrho F,\varrho G) .   \label{TCC}
\end{equation}
Here $\varrho^*$ is in principle any right inverse for $\varrho$, for~instance,
$$ \varrho^* f(\bq,\bp) := f(\znc)    $$
with the notations (\ref{nc-chng-vrbl-dgnrt-case}) and~(\ref{defZNC}).

\begin{remark} In~some sense, the freedom of choice for $\varrho^*$,
as~well as for the probability measure $d\mu$ in~(\ref{tilde-nu}),
reflects the ``degeneracy'' of the representation, as~does the reduction
of the number of variables of~$\eta$.
Note that the above choice for $\varrho^*$ has the virtue that it works
for all probability measures~$d\mu$ and values of $\hbar,\vthta$.
\end{remark}

Not~all choices of $d\mu$ and $\varrho^*$, however, are~physically relevant.
For the associated star-product
$$ F*G := \sum_{j=0}^\infty \hbar^j \tcC_j(F,G) , \qquad \text{i.e.} \quad
 \tcT_F\tcT_G\approx\tcT_{F*G}  ,   $$
we~would like to have the usual requirement that $\tcC_0(F,G)=FG$,
the~pointwise product. Applying $\varrho$ to~(\ref{TCC}), this implies
$$ \varrho(FG) = (\varrho F)(\varrho G)    \qquad \forall F,G .   $$
It~is easily seen that this is only possible when $d\mu$ is a Dirac~mass:
$$ d\mu(q_2,p_2) = \delta(q_2-q_2^*) \delta(p_2-p_2^*) ,  $$
for some fixed $(q_2^*,p_2^*)\in\RR^2$. The~functions $F,G$, being elements
of $L^\infty(\RR^4,d\tnu)$, are~then effectively defined only on the plane
$(q_2,p_2)=(q_2^*,p_2^*)$ (the~complement of this plane has zero measure);
and the right inverse $\varrho^*$ becomes simply the ordinary inverse. 
Viewing $F(\bq,\bp)=F(q_1,q_2^*,p_1,p_2^*)$ as a function of $q_1,p_1$ only,
and similarly for~$G$, we~then get as desired $\tcC_0(F,G)=FG$
(the~pointwise product), while
$$ \tcC_1(F,G) = -\frac1{2\hbar s^2}
  \Big(\hbar\frac{\partial F}{\partial p_1} -
       is^2\frac{\partial F}{\partial q_1}\Big)
  \Big(\hbar\frac{\partial G}{\partial p_1} +
       is^2\frac{\partial G}{\partial q_1}\Big) .   $$
Note that this expression does not depend at all on the parameter $\vthta$,
nor~on the choice of the base-point $(q_2^*,p_2^*)$; this~can be shown to
prevail also for $\tcC_j$, $j=2,3,\dots$.
In~fact, setting again $s=\sqrt{\hbar}$, we~arrive via (\ref{TCC}) just at
the formulas for the ordinary Berezin-Toeplitz quantization on~$\CC$ in the
``free'' variables~$q_1,p_1$.
 
\section{Conclusion and Outlook}\label{sec:conclsn}
In this paper, we have constructed noncommuative coherent states associated with a system of 2 degrees of freedom by means of the continuous families of UIRs of the kinematical symmetry group $\g$ of the underlying system. Subsequently, we computed the pertaining reproducing kernels. Since the generic families of UIRs are indexed by 3 nonzero continuous parameters $\hbar$, $\mathcal{B}$ and $\vartheta$ subject to a quadratic constraint $\hbar^{2}-\mathcal{B}\vartheta\neq 0$, it was natural to quest for the Berezin-Toeplitz quantization of the observables on the underlying 4-dimensional Phase space using all 3 deformation parameters instead of the single Planck's constant $\hbar$. In fact, when $\mathcal{B}$ (or $B=\frac{\mathcal{B}}{\hbar}$) and $\vartheta$ (or $T=\frac{\vartheta}{\hbar}$) are both zero, the UIRs of $\g$ indexed by this single nonzero Planck's constant $\hbar$ are nothing but the UIRs of the 2-dimensional Weyl-Heisenberg group $\GWH$. But the asymptotic analysis of Berezin-Toeplitz quantization pertaining to the generic sector (where $\hbar^{2}-\mathcal{B}\vartheta\neq 0$) of the unitary dual of $\g$ reveals the fact that $\hbar$, $\mathcal{B}$ and $\vartheta$ cannot all approach 0 independently due to the imposed quadratic constraint. But upon ``renormalizing'' the deformation parameters $\mathcal{B}$ and $\vartheta$ to $B$ and $T$, respectively, one achieves the desired semiclassical asymptotics with $B$, $T$ and $\hbar$ simultaneously approaching 0. 

We have subsequently handled the degenerate case $\hbar^{2}-\mathcal{B}\vartheta=0$ or $BT=1$ by defining the associated family of noncommutative coherent states on a dimensionally reduced Hilbert space and observed the crucial fact that the deformation parameter $\vartheta=\frac{\hbar^{2}}{\mathcal{B}}$ disappears from the picture completely yielding the standard setting of Berezin-Toeplitz quantization on the complex plane $\mathbb{C}$.
Our analysis of the degenerate noncommutative coherent states is propelled by the family of UIRs (see (\ref{eq:degenerate-rep-momentumspace-inv-trnsfrm})) of $\g$. In other words, the group representation theoretic analysis of NCQM conducted in \cite{ncqmjpa} has enabled us to study such degenerate setting in the context of NCQM. As has been insinuated towards the end of Section \ref{sec:intro} that $B$ here stands for the constant magnetic field applied perpendicular to a charged particle constrained to move on a two dimensional plane. The degenerate case that we have studied here is closely tied with the massless limit of a charged particle moving on such a plane subject to a vertical constant magnetic field (p. 213, \cite{Szabo}). Such a model has zero Hamiltonian in the limiting case turning the theory into a topological one. The string theoretic analog of this degenerate case is also discussed there in \cite{Szabo}.

In an earlier paper \cite{nctori} by one of the present authors, noncommutative
4~tori have been constructed explicitly using the unitary dual of $\g$.
We~propose to undertake an in-depth study of the pertinent aspects from the
point of view of noncommutative geometry, in~particular, classifying projective
modules over this NC-4~tori, computing connections with constant curvature and
the Chern character on the relevant projective modules in near future.

\end{document}